# INFLUENCE OF PRESSURE ON THE KINETICS OF FERROELECTRIC PHASE TRANSITION IN BaTiO$_3$

Olga Mazur[1,a)], Ken-ichi Tozaki[2], Yukio Yoshimura[3], Leonid Stefanovich[1]

[1] Institute for Physics of Mining Processes of the National Academy of Sciences of Ukraine, Symferopolska st., 2a, Dnipro-5, 49600, Ukraine

[2] Department of Physics, Faculty of Education, Chiba University, Chiba-shi, Chiba, 263-8522, Japan

[3] Faculty of Science and Engineering, Ritsumeikan University, Kusatsu, Shiga 525-8577, Japan

[a)]**Author to whom correspondence should be addressed:** o.yu.mazur@gmail.com, +380509142572

## ABSTRACT.

The ferroelectric phase transition in barium titanate under pressure was studied within the framework of Landau – Ginzburg theory using differential scanning calorimetry. An innovative method for high-sensitive thermal measurements under pressure was demonstrated. It was shown that the relaxation process proceeds nonmonotonically with the formation of intermediate short-lived phases. It was established that polydomain ordering is preferred but the monodomenization of the sample is possible under certain conditions. A narrowing of the temperature hysteresis with increasing pressure was revealed. A new tricritical point in barium titanate crystals characterizing the changing of the phase transition has been experimentally determined and theoretically confirmed.

## GRAPHICAL ABSTRACT

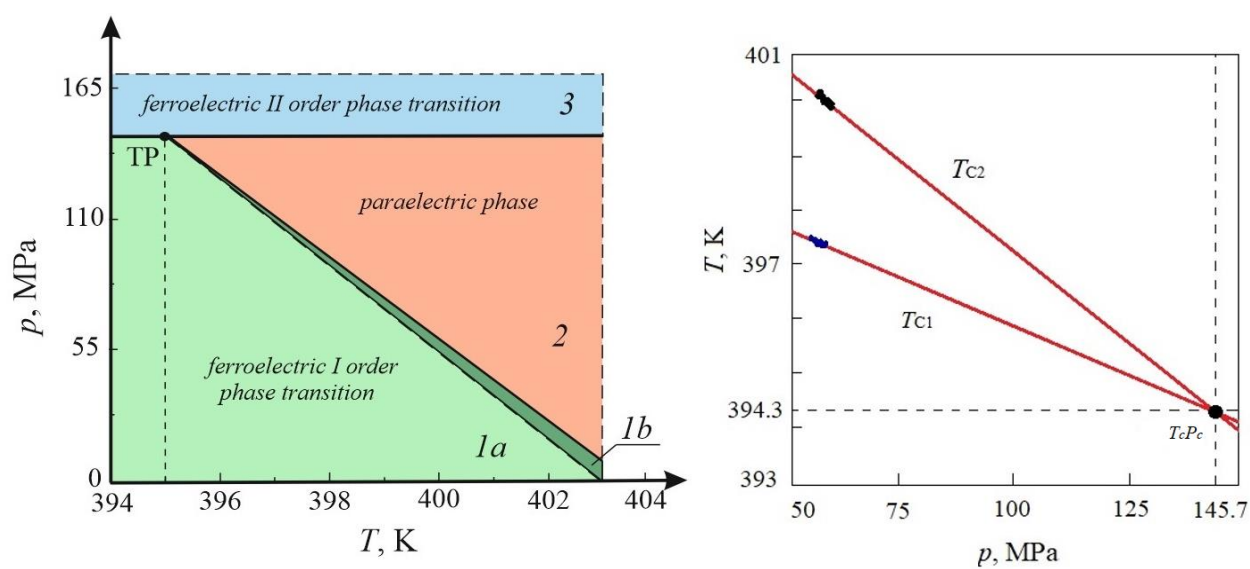

**Key Words: ferroelectric phase transition, barium titanate, tricritical point, differential scanning calorimetry, domain structure**

## 1. INTRODUCTION

Ferroelectric materials with the structure of perovskite $ABO_3$ remain one of the most studied objects of applied and fundamental physics of condensed matter. They have a wide range of unique properties and are especially in demand in modern instrumentation: energy and optical storage devices, waveguides, pyroelectric detectors and solar cells [1–3].

As is known, the physical properties of ferroelectric are determined by the state of its domain structure, which can change significantly under the influence of thermal, electrical and mechanical impacts. It was shown that the hydrostatic pressure can suppress the ferroelectric phase in perovskites and even destroy it [4].

The evolution of the domain structure to the equilibrium state after the transition to the polar phase is accompanied by changes in important macroscopic parameters of the ferroelectric. Crystals with the perovskite structure have large piezoelectric coefficients. Therefore, the use of hydrostatic pressure may become the most effective method to obtain domain structures of the required type in these materials. Strain engineering methods are already successfully used to improve the ferroelectric properties of perovskites [5, 6].

The thermodynamics of phase transitions in perovskites under the influence of hydrostatic pressure has been studied in considerable detail [7–11]. An explanation of the phenomenon of lowering the ordering temperature $T_C$ with increasing pressure using the concept of soft mode is given in [12]. In [13, 14] the tricritical behavior of barium titanate ($BaTiO_3$) and lead titanate ($PbTiO_3$) within the framework of phenomenological Landau theory was studied and the parameters of phase transition under the influence of mechanical stresses were calculated.

The use of high-pressure techniques allows for conducting experimental studies in the pressure range up to ~ $40 - 60$ GPa. Theoretical models already make it possible to extrapolate the obtained results and predict the behavior of the ferroelectric structure in the pressure range up to 200 GPa. Thus the stability of optical phonon modes in barium titanate at pressures up to 160 GPa was theoretically studied in [15]. Generalized ideas about the influence of external effects on the nature of the ferroelectric phase transition in perovskites are given in reviews [9, 16].

The technical difficulty of the observation of the "extended in time" phase transition especially under the influence of hydrostatic pressure imposed on the sample complicates experimental research. Many known experiments were performed, as a rule, on triglycine sulfate (TGS) crystals, which were quenched to the region below Curie temperature $T_C$ in the absence of external influences [17–20]. However, calorimetric studies of phase transition kinetics in barium titanate crystals have recently been actively conducted. These investigations show that phase transition occurs in a purely order-disorder-type two-step structural change at $T_C = 403$ K [21, 22].

Along with experimental research, the theoretical models of the kinetics of the formation of domain structures during the phase transition in ferroelectrics of different nature are being actively developed. Alternative phenomenological one- and two-dimensional self-consistent models which explain mechanisms of repolarization in disordered ferroelectrics were proposed in [23]. The influence of defects and doped impurities on the processes of ferroelectric domain formation in barium titanate crystal was studied in [24].

This work is devoted to the theoretical and experimental study of the kinetic aspects of ferroelectric phase transition in barium titanate crystal after its quenching in the vicinity of $T_C$ under the influence of mechanical stresses. Our research aims to establish the general kinetic regularities of the ferroelectric phase transition in $BaTiO_3$ and to develop practical recommendations for targeted control of domain structure formation.

## 2. PROBLEM STATEMENT

The appearance of a spontaneously polarized state during ferroelectric phase transition in oxides with a perovskite structure is explained by the Jhan-Teller cooperative pseudo effect. The ordered arrangement of dipoles is caused by distortions of nuclear configurations of the system in the minima of potential energy. In the paraelectric phase ($T > T_C$) these distortions have dynamic nature when there are continuous transitions between equivalent directions of distortion [25]. In the ferroelectric phase ($T < T_C$) nuclear distortions acquire a static character when the collective interaction of dynamically distorted complexes in the crystal volume leads to a structural phase transition of cubic symmetry crystals to the tetragonal phase [25].

According to the microscopic theory, the phenomenon of ferroelectricity in $BaTiO_3$ is related to the nature of the covalent Ti–O bonds and in particular to the charge coordination in $TiO_6$ octahedral clusters. Although this model makes it possible to elucidate the general regularities of phase transitions in ferroelectrics with a perovskite structure, the tasks of obtaining concrete results for direct comparison with experimental data are extremely difficult [7, 15, 25].

The Landau-Ginzburg-Devonshire phenomenological theory is widely used to describe relaxation processes in nonequilibrium systems [13, 14, 26]. This theory is successfully used by the authors of this paper to describe the kinetics of ordering in ferroelectric crystals that undergo phase transitions of the order-disorder type [27, 28]. The results obtained by analytical and numerical methods showed good qualitative agreement with the known experimental data [28, 29]. Therefore, it seems quite logical to further develop a phenomenological theory to describe the kinetics of the formation of domain structures in ferroelectrics of the displacive type.

Crystals with perovskite structures such as $BaTiO_3$ and $PbTiO_3$ have relatively simple crystal lattices and are often used both for experimental studies and as model objects for theoretical research. Despite this, many researchers obtain ambiguous results due to the use of different experimental conditions and theoretical models. Thus, according to the data from various sources, the ordering temperature of barium titanate $T_C$ varies in the range of 393–408 K [30–32]. Therefore, elucidation of the mechanisms of changes in the properties of crystals in different conditions remains an urgent task in the physics of ferroelectricity.

In this paper, the barium titanate crystal was chosen as an object for theoretical and experimental study. It undergoes the first-order ferroelectric phase transition from the cubic paraelectric phase into the tetragonal ferroelectric one [30–32].

The study of the formation of a domain structure during phase transitions may be carried out at pressures that are much lower than the value of the intracrystalline pressure of the sample. For this, the rapid quenching of the sample from the paraelectric phase into the ferroelectric one should be done. Cooling the sample into the temperature region below $T_C$ makes it nonequilibrium and very sensitive to even weak external influences. In this case, high pressures which can lead to the destruction of the crystal are not required at all.

Within the framework of the assigned task, the kinetics of the ferroelectric phase transition from the high-temperature high-symmetric paraelectric phase to the low-temperature ferroelectric phase under the influence of hydrostatic pressure is considered. As a result, regions with nonzero polarization spontaneously appear in different parts of the crystal — nuclei of ferroelectric domains which subsequently relax to a state of thermodynamic equilibrium. The change of external and relaxation conditions can significantly influence the rearrangement of domains at their primary formation.

The theoretical part of the work involves calculations of the control parameters (quenching temperature, hydrostatic pressure) to manage the state of the domain structure of quenched $BaTiO_3$ crystal and to predict the outcomes of the relaxation process. The experimental part of the work involves the study of the phase transition kinetics in $BaTiO_3$ using the differential scanning calorimetry (DSC) method with varying the value of hydrostatic pressure at a fixed temperature and vice versa.

### 3. THEORETICAL PART

#### 3.1. General description of the model.

To describe the quantitative changes in the system that undergoes phase transition one or more values called order parameter $\eta_i$ are introduced. In ferroelectric crystals, the order parameter is associated with the appearance of regions with a nonzero spontaneous polarization vector $\boldsymbol{P}_s$ in different parts of the crystal. The classical approach for studying the ferroelectric phase transition both in bulk and film perovskites is the Landau-Devonshire theory and representation of the Gibbs thermodynamic potential as an expansion of degrees of the order parameter [33]

$$G = \frac{1}{2}aP_z^2 + \frac{1}{4}bP_z^4 + \frac{1}{6}cP_z^6 + \frac{1}{8}dP_z^8 + ... \tag{1}$$

Here $P_z$ is the projection of the spontaneous polarization vector $\boldsymbol{P}_s$ on one of the crystallographic directions $z$ and parameters $a$, $b$, $c$, $d$ are coefficients of the expansion.

The description of ferroelectric phase transition is usually limited by taking into account only the terms $P_z^2$, $P_z^4$ for the second-order phase transition and $P_z^2$, $P_z^4$, $P_z^6$ for the first-order phase transition. But experiments show that for perovskite ferroelectrics it becomes necessary to take into account terms up to the eighth degree in polarization $P_z^8$ in the expansion of Gibbs potential (1). An estimation of the contribution of terms of different orders to the expansion (1) was represented in [34]. When the polarization $P_z$ is small (transition region close to paraelectric phase) the largest contribution to the free energy $G$ is made by the term of the order $P_z^4$. With the increase of the polarization $P_z$ (ferroelectric phase) the contribution of the term $P_z^8$ becomes significant. It should be noted that in the region of very low temperatures $T << T_C$ the difference between the contributions does not differ as much as near the Curie temperature $T \approx T_C$. Therefore, the term $P_z^8$ cannot be neglected when studying the influence of quenching temperature on the ordering kinetics in ferroelectrics under pressure. It was shown that in the case of writing the free energy in the form (1), where terms of the order of $P_z^8$ are taken into account, the model ideally describes the experimental data [9, 34].

Within the framework of the phenomenological theory the nonequilibrium thermodynamic Gibbs potential (1) can be written in the form of the Ginzburg – Landau functional [28]

$$\Phi_p = \int \left[ \left( \frac{1}{2}A(p,T)P_z^2 + \frac{1}{4}B(p)P_z^4 + \frac{1}{6}CP_z^6 + \frac{1}{8}DP_z^8 \right) + \frac{1}{2}\delta(\nabla P_z)^2 \right] dV . \tag{2}$$

Since it is supposed to investigate the kinetics of ordering under the influence of hydrostatic pressure, some parameters of functional (2) turn out to be functions of pressure. The main dependence of thermodynamic potential (2) on pressure $p$ is contained in two parameters – $A(p,T)$ and $B(p)$. The characteristic energy scale in this problem is of the order of the Curie temperature $T_C^{(0)}$ of the ferroelectric in the absence of pressure $p = 0$. Then other constants in functional (2) can be estimated as follows: $C$, $D \sim T_C^{(0)}$. For structural phase transitions even when the long-range dipole interaction dominates the coefficient $\delta$ can be written as $\delta \sim T_C^{(0)}\rho_0^2$ ($\rho_0$ is the radius of interatomic interaction), i.e. the correlation energy is not abnormally large.

Substitution of functional (2) into the Landau – Khalatnikov relaxation equation [27] allows for describing the evolution of a nonequilibrium system which gives the following equation

$$\frac{\partial \pi}{\partial \tau} = \Delta\pi + \alpha(p,T)\pi + \beta(p)\pi^3 - \pi^5 - \pi^7 .$$ (3)

Here $\Delta$ is the Laplace operator; $\pi = P_z/P_s$ is the dimensionless order parameter, $P_s$ is the saturation polarization on the hysteresis curve. Functions $\alpha(p,T)$ and $\beta(p)$ define parameters $A$ and $B$ dimensioned on the Curie temperature $T_C^{(0)}$ and describe the influence of pressure $p$ and quenching temperature $T$ on the domain structure formation. Time is assumed to be measured in units $t_i$ (time of the elementary act of the system restructuring) and the spatial scale is in units $\rho_0$. Equation (3) describes the space-time evolution of the order parameter under the influence of external hydrostatic pressure.

Parameter $\alpha(p,T) \sim (T-T_C(p))/T_C^{(0)}$ describes the proximity of the temperature to which the sample is quenched $T$ to the Curie temperature $T_C(p)$, which depends on the pressure [28]. Barium titanate has a soft mode that defines phase transformations in the crystal under external influences. The vanishing of the soft mode at the phase transition point reflects the tendency to zero the difference between the short-range and long-range Coulomb forces. The compression of the crystal under the hydrostatic pressure leads to the fact that the short-range interaction grows faster than the long-range one, as a result of which a shift of the Curie temperature $T_C$ occurs [9]. With an increase in hydrostatic pressure, the condensation temperature of the central soft mode in barium titanate decreases monotonically. From literary sources, it is known that the pressure coefficient of BaTiO$_3$ has a value $\partial T_C/\partial p = -55$ K/GPa [35].

Parameter $\beta(p) \sim (p/p_{tcr} - 1)$ characterizes the closeness of the hydrostatic pressure $p$ imposed on the sample to the tricritical value $p_{tcr}$. In a range up to the tricritical pressure $p < p_{tcr}$ parameter $B$ in functional (2) has a negative sign. When the value of pressure is close to the tricritical magnitude $p \rightarrow p_{tcr}$ the phase transition begins to acquire the features of a second-order phase transition. If the hydrostatic pressure imposed on the system reaches the tricritical value $p = p_{tcr}$ parameter $B(p)$ in functional (2) vanishes. With a further increase of pressure parameter $B(p)$ becomes positive and the system will already experience a second-order phase transition.

As a result of the rapid cooling of the system, the spatial fluctuations of the order parameter (polarization) spontaneously appear throughout the entire volume of the sample. Since they are randomly distributed in space the initial state of the ordering system should be set statistically. The rate of quenching is assumed to be slow enough so that the crystal does not crush, but fast enough so that the system does not instantly go into a thermodynamically stable state. Experiments show that the observation of nonequilibrium processes in BaTiO$_3$ both upon heating and cooling is possible with a quasi-static temperature change at a rate of 3µK/s [22]. In this case, the evolution of the domain structure can take from several minutes to several hours, depending on the relaxation conditions and the influence of external impacts.

To describe the relaxation of the system to the state of thermodynamic equilibrium it is enough to find the average value of the order parameter $\langle \pi(\mathbf{r},\tau) \rangle \equiv \bar{\pi}(\tau)$ and the two-point correlation function $K(s,\tau)$, where $\mathbf{s} = \mathbf{r}' - \mathbf{r}$ is the distance between two points $\mathbf{r}'$ and $\mathbf{r}$ in different domains respectively [27]. The procedure for obtaining equations for the average polarization $\bar{\pi}(\tau)$ and the second-order correlator for the 180º ferroelectric domain structure we developed in [27]. Due to its universal character it can be used for the equation (3) under certain restrictions. First of all, this is a study of the shallow quenching of the sample ($T_C - T \leq 10$ K), where the only tetragonal phase is realized. The second assumption is that the sample under consideration is sufficiently thin. Thus, as a result of quenching, the entire volume of the sample becomes nonequilibrium and a 180º domain structure is formed in it.

According to the procedure written in detail in [27, 28] the system of evolution equations for average polarization $\bar{\pi}$ and its dispersion $D$ was obtained

| Coefficients $k_{ij}$ | | | | |
|---|---|---|---|---|
| $i$ $j$ | 0 | 1 | 2 | 3 |
| 0 | $\alpha(p,T)$ | $\beta(p)$ | -1 | -1 |
| 1 | $3\,\beta(p)$ | -10 | -21 | 0 |
| 2 | -5 | -35 | 0 | 0 |
| 3 | -7 | 0 | 0 | 0 |

$$\begin{cases} \dfrac{d\overline{\pi}}{\partial t} = \dfrac{1}{2}\sum_{i=0}^{3}\sum_{j=0}^{3} k_{ij}\,\overline{\pi}^{2i+1} D^{j} \\[4mm] \dfrac{dD}{\partial t} = \sum_{i=0}^{3}\sum_{j=0}^{3} k_{ij} D^{i+1}\overline{\pi}^{2j} \end{cases} \quad (4)$$

Table 1. Quenching parameter α(p,T) is written in this form only in the first equation of the system (4). In the second equation it has the form $\alpha_{eff}(p,T,\tau)$ [28]. Function $\alpha_{eff}(p,T,\tau)$ defines the evolving character of domain structure in time. It depends on the quenching temperature and hydrostatic pressure imposed on the sample through the parameter α(p,T) and correlation radius $\rho_c(0)$ [28].

*Qualitative analysis*

The nonlinear character of the system of equations (4) determines many scenarios for the evolution of the domain structure to the state of thermodynamic equilibrium depending on the initial conditions. But due to the same reason, it cannot be solved analytically. Qualitative analysis of the system (4) can be made in some limiting cases:

- the system is quenched in the vicinity of the phase transition point and the initial dimensionless correlation radius $\rho_c(0)$ is small, i.e. the inequality holds $\rho_c(0) << 1/(\alpha)^{1/2} << d$, where $1/(\alpha)^{1/2}$ is the characteristic size of the domain; $d$ is the characteristic size of crystallite. All values are in units $\rho_0$.
- relaxation of the system is studied at large times $\tau >> \tau_d \sim 1/\alpha$, where $\tau_d$ is a characteristic time of domain structure formation.

In the case of asymptotic consideration of large times ($\tau \to \infty$) the system of equations (4) is greatly simplified. Parameter $\alpha_{eff}$ tends to the constant value $\alpha_{eff} = \alpha = const$ and left sides of equations (4) vanish $d\overline{\pi}/d\tau = 0$; $dD/d\tau = 0$. Then the system of differential equations (4) is reduced to a system of algebraic equations.

Qualitative analysis of the asymptotic system of equations using the phase pattern concept gives the information about final stages of the ordering process [28]. When the sample is quenched to the region below Curie temperature $T < T_C$ ($\alpha > 0$) the complex of singular points arises in the phase pattern (Fig. 1). Relaxation of the system was considered under the influence of hydrostatic pressure in the absence of an external electric field, which could shift the positions of singular points [27]. Therefore, the phase pattern is symmetric and only singular points located in the right quadrant of the phase pattern can be considered (Fig. 1). As a result of solving the system of equations (4) at large times, the following singular points were obtained

$$\begin{cases} (\text{I}): \quad \overline{\pi} = 0;\ D = 0 \\[3mm] (\text{II}): \quad \overline{\pi} = \beta^{1/2} + \dfrac{\alpha\beta^{1/2}}{2\beta^2(3\beta+1)};\ D = 0 \\[3mm] (\text{III}): \quad \overline{\pi} = 0;\ D = \beta + \dfrac{\alpha}{\beta(3\beta+1)} \\[3mm] (\text{IV}): \quad \overline{\pi} = \dfrac{1}{2}\left(\beta + \dfrac{\alpha}{\beta(3\beta+1)}\right)^{1/2};\ D = \dfrac{1}{4}\left(\beta + \dfrac{\alpha}{\beta(3\beta+1)}\right) \end{cases} \quad . \quad (5)$$

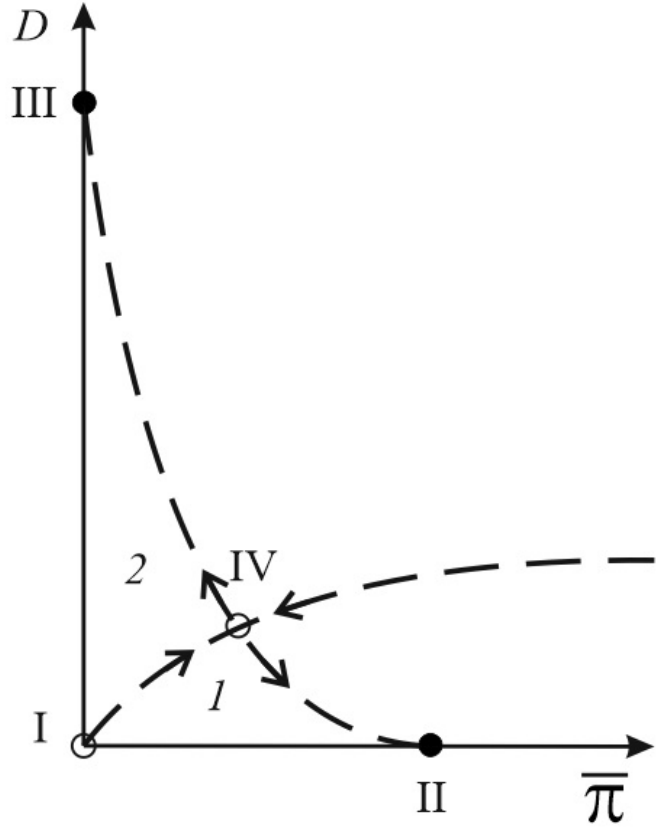

**Fig.1.** Qualitative phase pattern of the asymptotic system in variables $(\bar{\pi}, D)$. Singular points of the system are indicated by Roman numerals: black points correspond to stable states and white points correspond to unstable states. The dashed lines indicate the separatrices. Sectors 1 and 2 correspond to the regions of single-domain and polydomain structure formation respectively.

The type and form of singular points at different relations between parameters α and β are presented in detail in [28]. The singular point I is an unstable node that characterizes the initial state of the system before its quenching, i.e. in the paraelectric phase. Singular points II and III are stable nodes that correspond to the formation of stable single-domain and polydomain structures respectively. Singular point IV is a saddle point that defines the short-lived metastable asymmetric polydomain phase.

In this paper, the kinetics of ferroelectric phase transition is studied near phase transition temperature $T_C$ and points (5) are obtained for the general case. It should be noted that if the pressure is large enough and the effect of the quenching temperature is insignificant (α → 0), the singular points (5) take the form of the points obtained in [28].

Separatrices divide the phase pattern of the system into two sectors (Fig. 1). Sectors *1* and *2* are regions of attraction of the single-domain and polydomain ordering states respectively. Obviously, the larger the initial dispersion $D_0$, i.e. polarization inhomogeneity arising during quenching, the greater the probability of the evolution of the system to a stable polydomain structure. Similarly, the greater the value of initial polarization $\bar{\pi}_0$ the more domains of the same sign are formed at the initial stage of relaxation and the probability of the evolution of the system to a single-domain ordering state increases.

### 3.2. Numerical results of BaTiO$_3$

To describe the ordering kinetics of barium titanate crystal the numerical analysis of complete system of differential equations (4) was made using the following characteristic values of the sample: Curie temperature $T_C = 403$ K [21, 22], baric coefficient $\partial T_C / \partial p = -55$ K/GPa [35], Debye temperature $\Theta_D = 477$ K [36], activation energy $Q_{act} = 0.45$ eV [37]. Numerical analysis was carried out in the MatLab package by the classic Runge-Kutta method because it has high accuracy with a small number of calculations.

There were established three possible results of relaxation of nonequilibrium system near $T_C$ depending on the values of pressure $p$ and quenching temperature $T$: formation of stable domain structure (single-domain or polydomain) and the disappearance of the ferroelectric phase with a return to the paraelectric one.

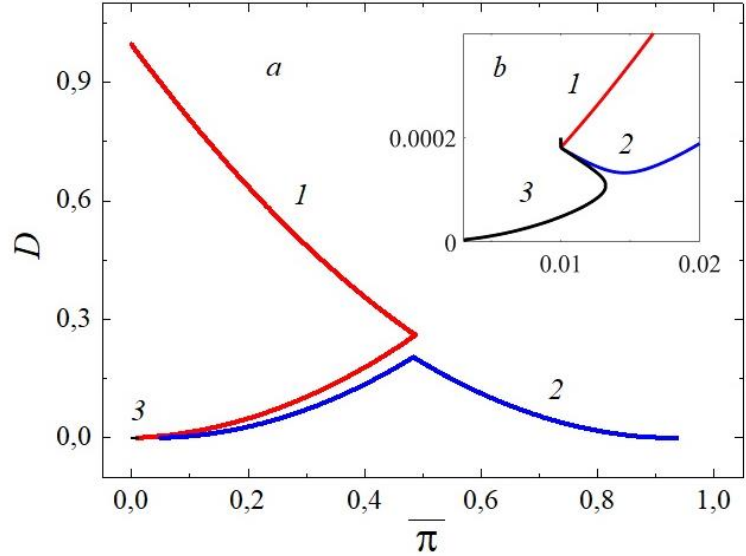

**Fig. 2.** Phase trajectories of the system, which describe the dynamics of the formation of the domain structure of BaTiO$_3$: $a$ – complete phase pattern; $b$ – a large-scale image of the initial relaxation phase of the system. Quenching temperature is $T = 402$ K. Quenching conditions in dimensionless units: initial average polarization $\bar{\pi}_0 = 0.01$, initial dispersion $D_0 = 0.0005$, initial average size of polarization inhomogeneity $r_0 = 1$. Phase trajectories $1 - 3$ correspond to the values of pressure 15, 23, 25 MPa respectively. Curves 1 and 2 show the appearance of a ferroelectric phase and the evolution of the domain structure to a stable polydomain and single-domain state respectively. Curve 3 reflected in the sidebar ($b$) characterizes the situation when the pressure suppresses the nuclei of ferroelectric domains formed after quenching, and the system returns to the disordered paraelectric phase.

The formation of a stable polydomain structure occurs at relatively weak pressure (curve 1 in Fig. 2). The growth of average polarization $\bar{\pi}$ indicates the gradual growth of domains of the one sign during the relaxation process. But if the pressure is small and the polarization inhomogeneities persist in the system then at a certain moment the active growth of domains of the opposite sign begins, which subsequently leads to the formation of a stable polydomain structure. Since the dispersion in the equilibrium state is $D \sim 0$ the formed domain structure turns out to be polydomain with equal volume fractions of domains of different signs. The degree of its inhomogeneity is determined by the value $D$.

Under specially selected initial conditions the relaxation process proceeds nonmonotonically accompanied by the presence of a metastable asymmetric phase near the saddle point IV (curves 1, 2 in Fig. 2). The duration of the kinetic slowing down of the system near the saddle point can be determined from the length of the step or plateau on the evolution curves (Fig. 3). The total time of polydomain structure formation in barium titanate crystal depends on the initial relaxation conditions. At quenching depth $T_C - T \leq 1$ K it can take up to 1 hour (curve 1 in Fig. 3) and the metastable state takes ¾ of this time. When the system overcomes this metastable phase it develops rather quickly and passes into a stable polydomain state (Fig. 3). But with the increasing of quenching depth both the total relaxation time and metastable phase time decrease. Thus at quenching depth $T_C - T = 7$ K it takes 480 seconds and 120 seconds respectively. A detailed study of the dependence of the relaxation time on pressure and quenching temperature is given in [28].

The existence of such metastable domain configurations that slowly move to an equilibrium state was predicted in [38, 39]. These processes can occur due to the motion of domain walls, changes in screening conditions, redistribution of defects, and other factors. Such kinks of polarization curves in BaTiO$_3$ (curves 1, 2 in Fig. 2) under pressure were experimentally studied in [40] and can be associated with the Barkhausen effect caused by the influence of mechanical stresses which was described in [28]. The appearance of such jumps of polarization near $T_C$ and their disappearance in the paraelectric phase was also observed in [41]. Pronounced Barkhausen jumps in a bulk BaTiO$_3$ were established during temperature changes close to the phase transition temperature and were explained by the domain structure instability in the ferroelectric phase close to the Curie point [41].

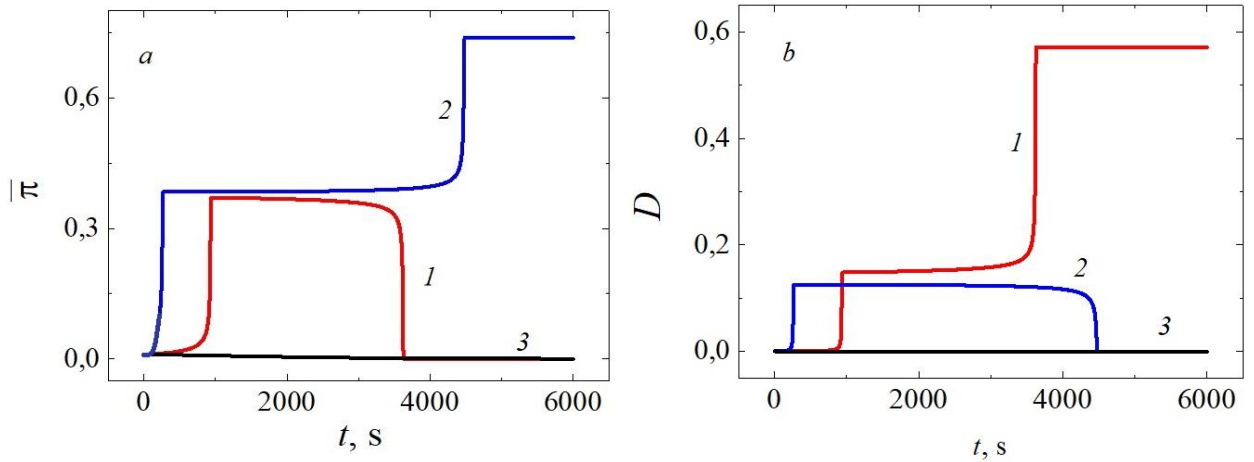

**Fig. 3.** Evolution of average polarization (a) and its dispersion (b) of ferroelectric BaTiO$_3$ under the same conditions as in Fig.2. Steps and plateaus on evolution curves 1 and 2 characterize the kinetic deceleration of the system near the saddle point and correspond to the polydomain states with the pronounced asymmetry of domains of the opposite sign of the polarization vector. Curve 3 shows an absence of the relaxation process after the suppression of the ferroelectric phase under pressure.

Ferroelectrics are often used as pyroelectric elements to create supersensitive uncooled temperature sensors and infrared detectors of thermal radiation. Therefore, the development of new methods for the formation of a stable single-domain structure in the sample is especially in demand. Polydomain structures with disoriented domains turn out to be unsuitable since in this case the pyroelectric effect is compensated. Monodomenization of perovskites in particular BaTiO$_3$ and PbTiO$_3$, is usually carried out by doping them with impurities during crystal growth. These impurities impede the movement of the domain walls. However, such an increase in the defectiveness of the crystal negatively affects its physical properties. Therefore, to create precision elements the alternative methods of monodomenization of samples are being developed.

The pyroelectric properties of ferroelectrics are determined by the primary pyroelectric effect. The pyroelectric coefficient reaches its temperature maximum and turns out to be most effective near $T_C$, where the change of spontaneous polarization with temperature becomes nonlinear. Therefore, the self-assembling formation of single-domain structure in the phase transition region under the influence of weak external effects can become a promising alternative method of pyroelectric elements development.

In the case of barium titanate, the use of hydrostatic pressure instead of an electric field will be more effective because this crystal also belongs to the class of ferroelastics, which are very sensitive to mechanical stress. In many studies, it was shown that the imposition of hydrostatic pressure on a BaTiO$_3$ crystal leads to the fact that, upon reaching a certain value, the total deformation reaches saturation and stabilization of the formed single-domain structure occurs [30]. But the spontaneous deformation in BaTiO$_3$ is the second-order effect and arises as a result of spontaneous polarization. Therefore, it is important to establish conditions at which the self-assembling monodomenization of the crystal will occur.

Our numerical results have indeed shown that with increasing pressure a stable single-domain structure can be formed (curve 2 in Fig. 2). The process of domain formation also proceeds nonmonotonically. However, with the correct selection of relaxation conditions, it ends with self-assembling monodomenization of the sample. The obtained evolution curves indicate that after the completion of relaxation the formed single-domain structure remains stable (Fig. 3). Experiments also show that the greater the pressure and the longer it is imposed on the sample the more domains do not return to their original state after the stress is removed [30]. Therefore, the influence of a weak pressure on the nonequilibrium system at phase transition can become a more promising method compared to the addition of impurities and an increase in the defectiveness of the crystal.

Evolution curves show that the relaxation of the system into a single-domain state proceeds longer than in the case of polydomain structure formation (curve 2 in Fig. 3). But the metastable asymmetric phase becomes faster and the slowdown of the system near saddle point IV (Fig. 1) occurs longer (Fig. 3).

The obtained results indicate that the domain structure in barium titanate forms relatively quickly. But in the vicinity of $T_C$ under some conditions, the long periods of stagnation (kinetic slowing down) when domain rearrangements occur very slowly can be observed. These asymmetric phases are characterized by increased internal energy in the system which takes some time to compensate. Similar to the case of polydomain ordering these processes are significantly accelerated with an increase in quenching depth and pressure.

It is important to note that large pressure leads to the destruction of the ferroelectric state, i.e. to the transition of the system into the paraelectric phase (curve 3 in Fig. 2). A change of the phase state under the influence of mechanical stresses, along with orientational and distortional effects of the domain structure, was noted in [30]. The nonmonotonic nature of the transition of the system to the ferroelectric phase is also indicated by the presence of an incubation period which is manifested by a slight decrease of dispersion at the early relaxation stage (curves 2, 3 in Fig. 2(b)). Indeed, it has been experimentally shown that the kinetics of domain formation in barium titanate is characterized by a stage of "latent" nucleation [30]. Moreover, the relaxation time of the system in this state linearly depends on the value inverse to the applied local load [28, 30].

As a result of numerical analysis two critical pressure values were determined: monodomain pressure $p_{mon}$ and paraelectric pressure $p_{par}$:

- $p < p_{mon}$ – a stable polydomain structure forms in the system;
- $p_{mon} \leq p < p_{par}$ – a stable single-domain structure forms in the system;
- $p \geq p_{par}$ – ferroelectric phase is suppressed and the system returns to a disordered state.

To establish the conditions for monodomenization of the sample within the framework of a numerical study the values of critical pressures $p_{mon}$ and $p_{par}$ were calculated at different quenching temperatures $T$, and a phase diagram was constructed (Fig. 4). It was established that with the increase of quenching depth both values $p_{mon}$ and $p_{par}$ increase.

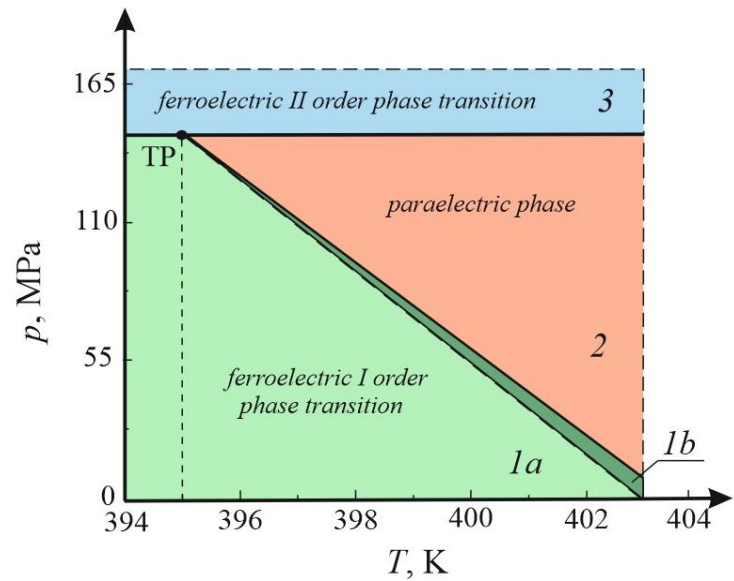

**Fig. 4.** Phase diagram $p - T$ with an indication of the tricritical point TP with coordinates $p = 145$ MPa, $T = 395$ K. In sector *1*, a first-order ferroelectric phase transition takes place: sector 1a corresponds to the polydomain structure formation; sector 1b corresponds to the single-domain structure formation. Sector *2* defines the region of the disordered (paraelectric) phase under pressure. Sector *3* corresponds to second-order ferroelectric phase transitions.

Interestingly, the range of control parameters for the self-assembling monodomenization of the sample is very tight (sector 1(b) in Fig. 4) and it narrows very quickly with the increase of quenching depth. Thus, when the sample is quenched in the vicinity of Curie temperature $T_C - T \leq 1$ K the monodomenization of the sample occurs in the pressure range $p = 19 - 24$ MPa. But when the quenching depth is $T_C - T = 2$ K this range is $p = 38 - 41$ MPa. And in the case of deep quenching $T_C - T = 7$ K the single-domain structure formation occurs only at pressure $p = 128$ MPa (Fig. 4).

The narrowing of the sector of monodomenization can be explained by the increase of critical stresses in a barium titanate crystal with the increase of the distance from Curie temperature [30]. Rapid cooling of the crystal leads to a simplification of its domain structure, as a reflection of internal mechanical stresses. Therefore, with an increase in quenching depth, the energetically more favorable polydomain ordering is observed predominantly. The gradual narrowing and complete disappearance of a single-domain phase at $T = 395$ K (Fig. 4) indicate a change in the nature of the phase transition from the first-order to the second-order where the polydomain structure formation is preferred [42].

Our results indicate that the point TP with coordinates $p = 145$ MPa; $T = 395$ K has the character of tricritical point (Fig. 4). However, this theoretical result requires experimental confirmation since the previously known tricritical points have different values $p = 3.4$ GPa; $T = 291$ K [43], $T = 323$ K [31] or $p = 6.5$ GPa; $T = 130$ K [32].

The developed theoretical model allows to determine control values of pressure $p$ and temperature $T$ at which ferroelectric (polydomain and single-domain state) and paraelectric phase can be realized. Numerical results show that the influence of pressure after the rapid cooling of the barium titanate crystal can be used for monodomenization of the sample, but only under weak pressure during quenching near the Curie temperature $T_C$. Thus, the developed model can be used to calculate control parameters and to direct the process of domain formation to obtain highly efficient pyroelectric elements.

## 4. EXPERIMENTAL PART

During cooling of the barium titanate single crystal from the high-symmetric region into the low-symmetric one, the first-order phase transition at $T_C = 403$ K ($\alpha$-transition) from the cubic paraelectric phase into the ferroelectric one is observed. After the cooling, the crystal is constructed by the tetragonal and monoclinic structures forming the coherent hybrid structure (CHS) [22]. Although the transition at $T_C$ is the first order, it also has the characteristics of a second-order phase transition in the vicinity of $T_C$ which is shown in dielectric dispersion [41, 44], an anomalous increase of specific heat [45, 46], softening of the crystals [47–49], and so on. When the temperature scanning speed is slowed down, the transition separates into two transitions, and a metastable intermediate state is observed between them [21, 22]. In other words, when the temperature of BaTiO$_3$ is brought closer to $T_C$ from the higher/lower side away from $T_C$ the fluctuation increases like in the case of second-order phase transition. However, the approach is interrupted at $T_{C1}/T_{C2}$ by the abrupt transition to the other phase accompanying a pulse-like exotherm/endotherm [22].

On the cooling/warming, the transition occurs irreversibly from the supercooled metastable state of the higher/lower temperature phase to the lower/higher temperature side phase at $T_{C1}/T_{C2}$. Since $T_{C1} < T_{C2}$ hysteresis is observed in the cycle of cooling and warming. Temperature range of the hysteresis $\Delta T = T_{C2} - T_{C1}$ depends on the height of the barrier between the two phases. When the barrier weakens, $\Delta T$ decreases. The main goal of the experimental part of the research was to study the dynamics of the phase transition under pressure and to establish how it affects the temperature hysteresis in BaTiO$_3$ crystal.

### 4.1. Methodology and Apparatus

Differential scanning calorimetry is one of the widely used methods of measuring of thermophysical properties of materials. Its main advantages include the speed of experiments, as well as high measurement accuracy at different rates of external temperature change. At the moment this is the simplest method for studying the phase state and kinetic analysis of a relaxing system. It allows not only to establish with high accuracy the type and temperature of the phase transition but also to determine the regions where temperature variation does not lead to a change in the phase state of the system. Kinetic analysis of the experimental DSC curves makes it possible to establish the reliability of the mathematical model used to calculate the main parameters characterizing the phase transition in the system. Thus, the DSC method is an indispensable tool in the study of critical phenomena in various materials and the construction of phase diagrams.

Experimental studies of the phase transition kinetics under pressure are associated with some technological difficulties. Therefore, many experiments were carried out either in the absence of external influences when only the quenching depth of the sample changed [17, 18] or under the influence of a weak electric field [50]. Some attempts of phase transition study under pressure have been made in the past, however, the detection sensitivity of conventional DSC is insufficient for a detailed investigation [51, 52]. Measuring capabilities of most commercial high-pressure DSC devices are limited by the pressure range up to 10–40 MPa only. But in the offshore industry, for example, pressures of 200 MPa and temperatures of 200 K are routinely encountered [53].

The DSC system used in this measurement was described in [54]. A schematic diagram of DSC device measuring the calorific value of a sample is shown in Fig. 5. A thermal module is glued to the upper and lower surfaces of the copper block to detect thermal anomalies in the sample. The sample is installed using a wire passed through a pipe attached to the thermal module. A platinum resistance thermometer (TS1) is included to measure the temperature of a portion of the sample. A porous thermal insulation material is used as the thermal insulation material, so that the working substance (silicone oil) is absorbed into the thermal insulation material. The pressurized medium is filled into a high-pressure container and the pressurization and depressurization are carried out using a high-pressure pump (SITEC 750,1200) controlled by a PC. A PC independently performs temperature and pressure control.

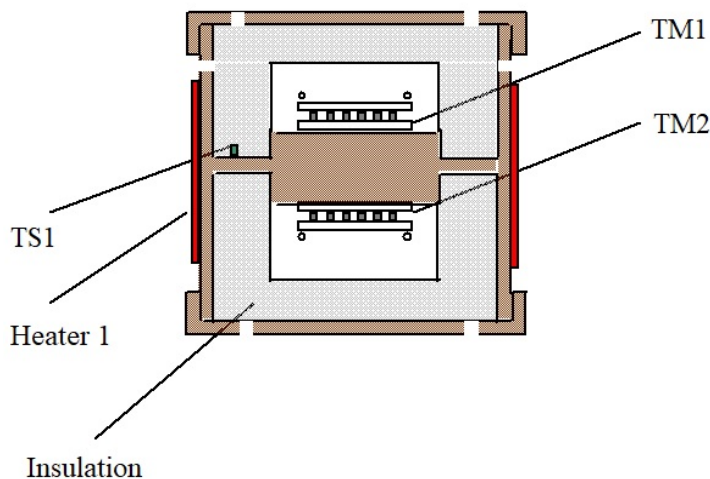

**Fig. 5.** Schematic of DSC unit. Two Peltier modules (TM1 and TM2) are glued on the upper and lower surfaces of the copper plate, respectively as the heat flow sensors. The sample is placed on the upper substrate of TM1. A platinum resistance thermometer (TS1) is included to measure the sample temperature. A porous ceramic was used as a thermal insulation material, so that the pressuring medium (silicone oil) to penetrate.

Below, we describe a DSC unit and a pressure vessel portion in which the DSC unit is installed. A schematic diagram of this DSC unit placed in a pressure container is shown in Fig. 6. It is equipped with a heater wrapped around the DSC unit for temperature control and a heater wrapped around the outer circumference of the pressure container. A high-pressure container is placed in a Dewar's vessel and its temperature is controlled using the second thermometer (TS2) and Heater 2 (Fig. 6).

The block diagram in Fig. 7 shows the state of the pressure regulation of the pump and the general control. The entire system is controlled by one PC and the values of temperature, heat and pressure are calculated by multiplying the data obtained in the program by the constants of each device. The pressure is monitored by a pressure transducer attached to the pump outlet and the pump motor is driven to achieve the required pressure.

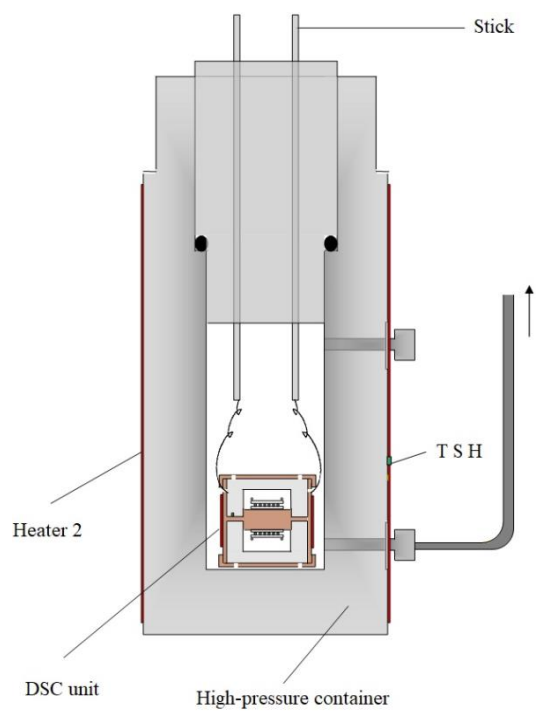

**Fig. 6.** Schematic of a high-pressure container with the DSC unit. A high-pressure container is placed in a Dewar's vessel and its temperature is controlled using the second thermometer (TS2) and Heater 2.

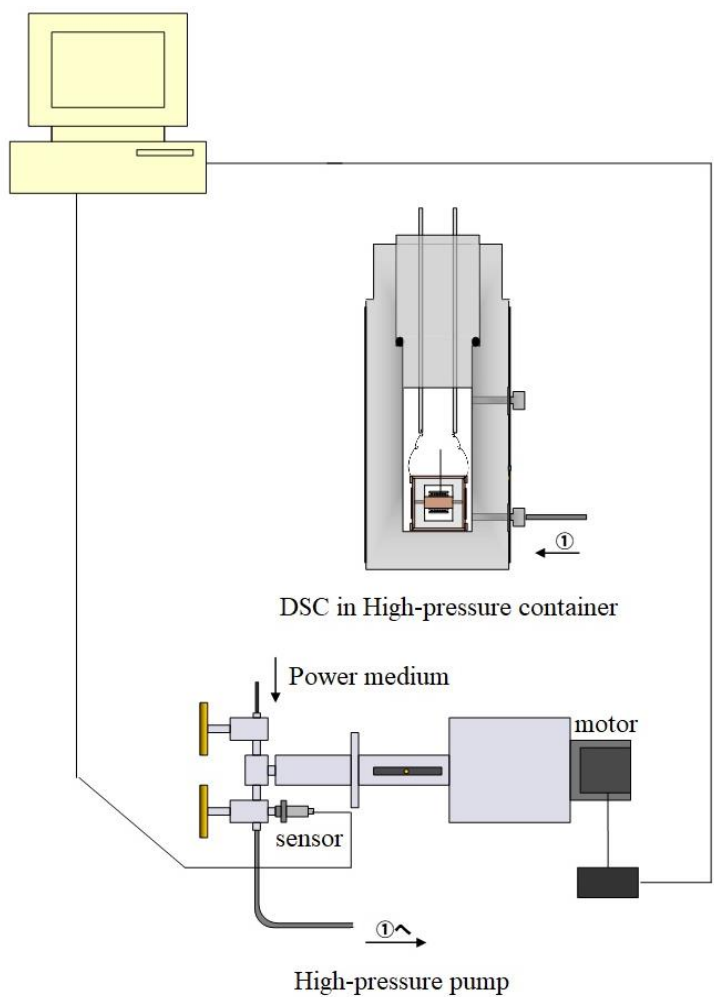

DSC in High-pressure container

Power medium

motor

sensor

High-pressure pump

**Fig. 7.** Block diagram of the pump pressure control system and general control.

### 4.2. Precise Measurements of Phase Transition in BaTiO₃ under pressure

The developed DSC allows us to make thermal measurements in a temperature range of 300 – 420 K and a pressure range of 0.1 – 200 MPa. The stability of temperature and pressure is ±1 mK and ±5 kPa, respectively. The detectable heat flow is 0.2 µW, which is less than 2 nW performed by highly stabilized DSC [54] under ambient pressure; however, it is proved to be sensitive enough to use for the study of the phase transition of BaTiO₃.

Measurements of temperature fluctuations at constant pressure (d$T$/dt = 0.5 mK/s) and measurements of pressure changes at constant temperature (d$p$/dt = 0.1 kPa/s) were made using BaTiO₃ solid crystal. A single crystal of BaTiO₃ was prepared by a top-seeded solution growth technique described in [58]. The single crystals were grown from the melt of the mixture of TiO₂=BaO 65/35 in molar ratio. The sources of TiO₂ and BaCO₃ were 99.99% and 99.999% purity powders respectively. The as-grown crystal boules with a dimension of about 20×20×15 mm$^3$ were cut to obtain a size of about 4×4×1.2 mm$^3$ [55].

### 4.3. Results.

It is known that at $T_C$ = 403 K the barium titanate crystal undergoes the first-order ferroelectric phase transition from the cubic phase into the tetragonal one at normal pressure. However, it shows sudden and multistep transformations when the temperature varies very slowly. This complexity may be affected by the surface of the specimen crystal.

The ferroelectric phase transition was studied under cooling and under heating of the sample at different values of pressure and quenching temperature. The quenching condition in theory is attained practically by non-deep super-cooling/heating in the experiment. These measurements allowed us to observe how the phase state of the nonequilibrium system changes depending on external influences.

At fixed pressure $p = 58.2$ MPa the temperature hysteresis that characterizes the first-order phase transition was observed (Fig. 8). It was found that the quenched paraelectric crystal transfers to a low-temperature ferroelectric phase abruptly on cooling at temperature $T_{C1} = 398.1$ K. But the transformation of the ferroelectric phase at superheated condition into paraelectric phase on warming occurs at temperature $T_{C2} = 400$ K. The temperature span $\Delta T$ observed as hysteresis on a cooling-heating cycle is given by $T_{C2} - T_{C1}$ and indicates the difficulty of the mutual transformation between both phases.

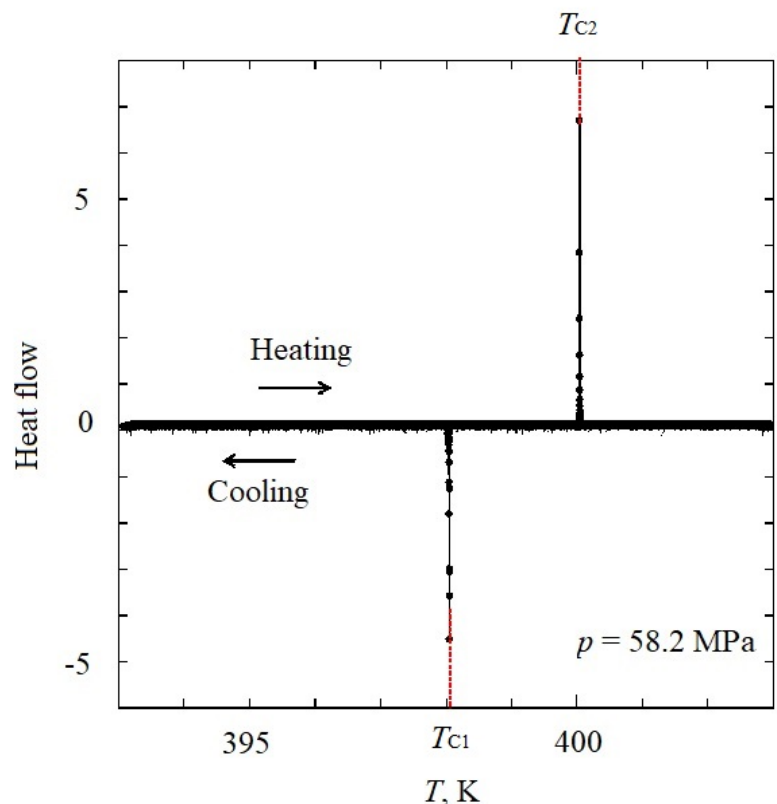

**Fig. 8.** Heat flow upon both cooling and warming at a rate of 0.5 mK/s of BaTiO$_3$ at fixed pressure $p = 58.2$ MPa.

Another task was to establish how this temperature hysteresis depends on pressure. Therefore, the phase state of the system was studied at some range of pressure values. Phase transitions were observed both at the increasing and decreasing pressure. It was established that with the growth of pressure both temperatures $T_{C1}$ and $T_{C2}$ decrease (Fig. 9a). The higher the pressure value, the smaller the difference $\Delta T$, which indicates a decrease in temperature hysteresis. The decreasing slope of $T$ by increasing the pressure is about –0.05 K/MPa, which value agrees with the results in [56].

The extrapolation of two regression lines for $T_{C1}$ and $T_{C2}$ shows the intersection at coordinates $p_1 = 145.7$ MPa, $T_1 = 394.3$ K (point $T_cP_c$ in Fig. 9b). At the point $(p_1, T_1)$ paraelectric and ferroelectric phases transform mutually by thermal fluctuations only and the phase transition occurs without the hysteresis phenomenon. Therefore, the point $T_cP_c$ can be defined as a tricritical point where the first-order phase transition transforms into the second-order phase transition. It can be noted that decreasing of temperature hysteresis obtained experimentally (Fig. 9b) has the same form as the decreasing of a single-domain region formation on a phase diagram obtained theoretically (Fig. 4). In our opinion, the point $T_cP_c$ can be associated with the theoretically calculated point TP with coordinates $p = 145$ MPa; $T = 395$ K (Fig. 4). These results allow us to conclude that the implementation of a single-domain structure after quenching the sample below $T_C$ is a distinctive feature of a first-order phase transition.

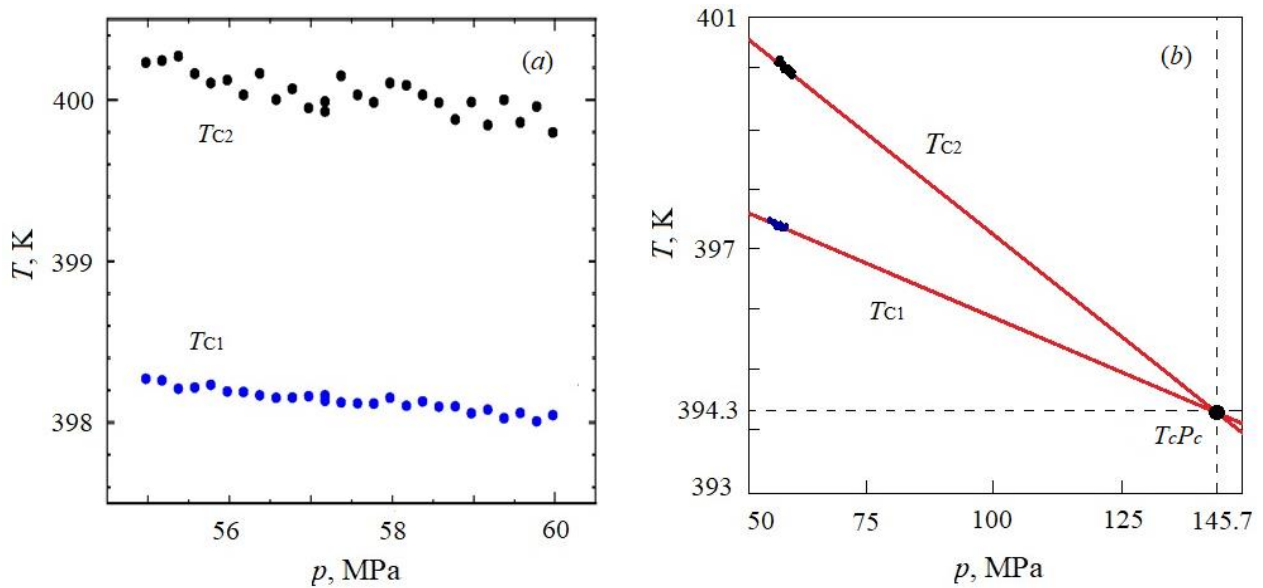

**Fig. 9.** (a) - temperature hysteresis of $BaTiO_3$ depending on pressure and quenching temperature values; (b) - extrapolation of regression of phase transition temperatures $T_{C1}$ and $T_{C2}$ depending on pressure and quenching temperature with the indicating of tricritical point $T_cP_c$.

The results of our study show good agreement between theoretical calculations and experimental measurements. Thus, the DSC method can be very useful for high-sensitive thermal-pressure studies of ferroelectric samples and the developed theoretical model can be applicable for preparing such experiments.

## 5. DISCUSSIONS.

Interest in the study of tricritical phenomena in ferroelectrics is not only fundamental but also applied goals. Although the cubic and tetragonal phases have the same free energy for the first-order phase transition, there is an energy barrier between the two phases [57]. The disappearance of this barrier near the tricritical point leads to a large polarization change. As a result, the material becomes easily polarized with a large dielectric permittivity and high energy density at a low electric field. Therefore, investigations of the tricritical phenomena in ferroelectrics can be very beneficial for the development of low voltage energy storage devices [57].

The tricritical points previously found in barium titanate are far from the Curie temperature ($T_C - T > 80$ K) and are characterized by high pressure ($p \sim 3$–6 GPa) [31, 32, 43]. When studying the kinetics of the phase transition near the $T_C$, we found a tricritical point at a relatively low-pressure $p \sim 145$ MPa. This tricritical behavior can be explained by the fact that in the phase transition region the quenched ferroelectric is very sensitive to any external influences. And even very weak pressure can significantly change the dynamics of a nonequilibrium system. Such phenomenon was observed previously in other ferroelectrics. For example, quasi-tricritical behavior and minimal thermal hysteresis ($\Delta T \approx 0.05$ K) were observed in the vicinity of $T_C$ in TGSe crystal [58, 59].

We did not find data about tricritical points near $T_C$ in barium titanate crystals. But the multicritical behavior in this ferroelectric at phase transition region was predicted in many works [44, 60, 61] and can be confirmed by theoretical and experimental results [57]. It follows from Landau's theory that in the tricritical region the dielectric permittivity increases sharply and shows peak values at the tricritical point. The clear BaTiO$_3$ sample has not so large dielectric permittivity as barium titanate composites, but it shows some peak at temperature $T \approx 395$ K which indicates the tricritical behavior of the system in this region [57] and is in good agreement with our theoretically and experimentally obtained tricritical point TP – $T_c P_c$.

Theoretical calculations of tricritical points obtained by Landau-type modeling correlate well with experimental data for many ferroelectrics, including perovskites: KDP [28, 35, 62], TGSe [59], MAPbI$_3$ [63], Sn$_2$P$_2$S$_6$ [64], DMAGaS [65], BaTiO$_3$ [44, 57, 60, 61]. The existence of tricritical points and changing of phase transition dynamics are associated with the relation between pressure and temperature values. In displacive and hydrogen-bonded ferroelectrics the Curie temperature shifts generally downwards with pressure. The disappearance of the ferroelectric phase under pressure was predicted before and is the most pronounced in ferroelectrics with a quite high baric coefficient $\partial T_C / \partial p$ [35].

Within the framework of our kinetic theory the tricritical behavior was established for ferroelectrics KDP, (NH$_4$)$_2$SO$_4$ and BaTiO$_3$ with the negative sign of baric coefficient ($\partial T_C / \partial p < 0$) since for ferroelectrics KNO$_3$, NH$_4$HSO$_4$ and TGS with positive sign of baric coefficient ($\partial T_C / \partial p > 0$) it was not observed [27, 28, 42]. A similar situation is in dimethylammonium sulfate ferroelectrics: DMAAS ($\partial T_C / \partial p > 0$) does not have a tricritical point and DMAGaS ($\partial T_C / \partial p < 0$) has it [65, 66]. These results suggest that at temperature region $T < T_C^{(0)}$ the tricritical behavior under pressure is observed only in ferroelectrics with a positive sign of the compressibility coefficient, i.e. when the Curie temperature decreases with increasing pressure [42]. Indeed, in crystals KNO$_3$ [67], TGSe [59], NH$_4$HSO$_4$ [35] the shift of Curie temperature with the pressure led to the quasi-critical and tricritical behavior at temperatures $T > T_C^{(0)}$.

The expansion of the thermodynamic potential within the Landau theory to the sixth degree of the order parameter describes well the tricritical behavior in many perovskites. However, Landau's theory also predicts a vanishing region of phase coexistence at the tricritical point, which is inconsistent with the observations in [58, 63]. A possible reason for this mismatch is the influence of critical fluctuations [46] or the importance of taking into account higher-order terms in the expansion (1) [34, 58, 63]. Landau's theory neglects the influence of fluctuations, so we cannot verify this reason within the framework of this research, but we studied the second assumption. We carried out the numerical calculations within the framework of both models with the expansion of the functional up to the eighth degree and up to the sixth degree of the order parameter. In the case of relatively weak pressure (up to 200 MPa) even in the vicinity of $T_C$ the corrections introduced by the eighth degree of the order parameter practically do not change results. Therefore, despite the conclusions made in [34, 58, 63] the consideration of ferroelectric phase transition under pressures up to 200 MPa can be made within the theoretical model with the expansion of the thermodynamic functional to the sixth degree of the order parameter presented in [28].

The kinetic theory we developed predicts the tricritical behavior of the system regarding the evolving character of domain structure formation. The disappearance of a single-domain state with an increase in pressure and quenching temperature indicates a change in the type of phase transition from the first-order to the second-order where the polydomain state is preferred [42]. The presence of a bulk disorder which is typical for perfect samples induces the changing of the dynamic phase transitions and the appearance of different self-assembling criticalities [44]. Therefore, it is very important to use write parameters for numerical modeling for the theoretical description of the kinetic processes near $T_C$.

The good agreement of theoretical results with the experiment is due to the qualitative selection of simulation parameters and initial conditions (indicated in figure caption for Fig. 2). In this numerical study, within the framework of the constructed model, a nominally ideal sample was assumed, which has a phase transition temperature $T_C = 403$ K and a baric coefficient $\gamma = -55$ K/GPa [35]. If in the experiment a crystal grown under special conditions is selected, then its initial parameters, $D_0$, $\rho_0$, as well as the Curie temperature $T_C$ would change. Due to the connection between the ordering temperature and the pressure value through the baric coefficient $\partial T_C / \partial p$ the set critical point will shift too.

In addition to the experimental verification of the theoretical calculations obtained in this work, their reliability is confirmed by good agreement with observations in [30, 41, 44, 57]. This correlation of the calculated results with the data obtained in the experiments shows the applicability of the developed model for describing the main characteristics of the phase transition for a certain ferroelectric crystal. This makes it possible to predict the states of the domain structure depending on its type and values of control parameters.

## 6. CONCLUSION.

Kinetics of ferroelectric phase transition in $BaTiO_3$ under the influence of pressure was theoretically and experimentally studied within the framework of phenomenological Landau – Ginzburg – Devonshire theory and differential scanning calorimetry method.

The theoretical study showed that domain structure formation can proceed nonmonotonically with the formation of metastable asymmetric polydomain phases and incubation period at the early stage of relaxation. Depending on the values of pressure and quenching temperature three outcomes of evolution can be observed: formation of stable ferroelectric domain structure (polydomain or single-domain) or return of the system into disordered paraelectric phase. It was shown that the quenched system tends to form a stable polydomain state. The increase of pressure puts the system into a single-domain state but the further increase of this value turns it into the paraelectric phase.

The proposed method of pressure imposing on the perovskite ferroelectric that undergoes phase transition can be used for monodomenization of the sample. The developed model allows us to calculate values of pressure and quenching temperature at which the thermodynamically stable single-domain structure forms. Two critical pressures were defined: $p_{mon}$ which suppresses the inhomogeneity and forms a stable single-domain state and $p_{par}$ which destroys the ferroelectric ordering at all.

It was shown that with an increase in quenching depth, the range of pressure values at which a polydomain structure is formed expands. In this case, the range of values for the formation of a single-domain structure narrows and completely disappears at the point $p = 145$ MPa; $T = 395$ K. This behavior of the system allows us to classify this point as the tricritical one obtained for the first time.

An experimental study of the ferroelectric phase transition was carried out using differential scanning calorimetry. The laboratory-made DSC allows for providing high-sensitive measurements of physical properties with the changing of temperature and pressure values.

The temperature hysteresis was established and the values of the phase transition temperature during cooling and heating of the sample were determined. It was shown that with the increase of pressure the temperature hysteresis decreases and disappears at an extrapolated point of $p_1 = 145.7$ MPa, $T_1 = 394.3$ K, which can be classified as a tricritical one. Thus, a good agreement was established between theoretical and experimental results.

The experiment performed shows that the imposing of pressure on the sample during its quenching from the paraelectric phase into the ferroelectric one can be used in practice to control the physical properties of the perovskite material. The developed theoretical model makes it possible to calculate the control parameters for managing the state of the emerging domain structure. It is expected that the results of our study will find a practical application and will be used to obtain ferroelectric samples with specified parameters.

## ACKNOWLEDGMENTS

Authors thank Mr. O. Nakao of Materials Research Laboratory of Fujikura Ltd., for providing the single crystal of BaTiO$_3$.

## DECLARATION OF INTERESTS

The authors declare that they have no known competing financial interests or personal relationships that could have appeared to influence the work reported in this paper.

## DATA AVAILABILITY

The data that support the findings of this study are available from the corresponding author upon reasonable request.